\documentclass[preprint,amsmath,amssymb,prb,floatfix,superscriptaddress]{revtex4}
\usepackage[dvips]{graphicx}% Include figure files
\usepackage{dcolumn}% Align table columns on decimal point
\usepackage{bm}% bold math
\usepackage{mathrsfs}

\begin{document}

\title{Thin InSb layers with metallic gratings: a novel platform for spectrally-selective THz plasmonic sensing}

\author{Shuai Lin}
 \affiliation{Department of Physics and Engineering Physics, Tulane University, 6400 Freret St., New Orleans, LA 70118, USA}
\author{Khagendra Bhattarai}
 \affiliation{Department of Physics, The University of South Florida, Tampa, Florida 33620-7100, USA}
\author{Jiangfeng Zhou}
 \email{jiangfengz@usf.edu}
 \affiliation{Department of Physics, The University of South Florida, Tampa, Florida 33620-7100, USA}
\author{Diyar Talbayev}
 \email{dtalbaye@tulane.edu}
 \affiliation{Department of Physics and Engineering Physics, Tulane University, 6400 Freret St., New Orleans, LA 70118, USA}
 
\date{\today}

\newcommand{\cm}{\:\mathrm{cm}^{-1}}
\newcommand{\T}{\:\mathrm{T}}
\newcommand{\mc}{\:\mu\mathrm{m}}
\newcommand{\ve}{\varepsilon}
\newcommand{\dg}{^\mathtt{o}}

\begin{abstract}
We present a computational study of terahertz optical properties of a grating-coupled plasmonic structure based on micrometer-thin InSb layers.  We find two strong absorption resonances that we interpret as standing surface plasmon modes and investigate their dispersion relations, dependence on InSb thickness, and the spatial distribution of the electric field.  The observed surface plasmon modes are well described by a simple theory of the air/InSb/air trilayer.  The plasmonic response of the grating/InSb structure is highly sensitive to the dielectric environment and the presence of an analyte (e.g., lactose) at the InSb interface, which is promising for terahertz plasmonic sensor applications.  We determine the sensor sensitivity to be 7200 nm per refractive index unit (or 0.06 THz per refractive index unit).  The lower surface plasmon mode also exhibits a splitting when tuned in resonance with the vibrational mode of lactose at 1.37 THz.  We propose that such interaction between surface plasmon and vibrational modes can be used as the basis for a new sensing modality that allows the detection of terahertz vibrational fingerprints of an analyte.
\end{abstract}

\maketitle

%%%%%%%%%%%%%%%%%%%%%%%%%%  body  %%%%%%%%%%%%%%%%%%%%%%%%%%
\section{Introduction}
Surface plasmon polaritons (SPPs) are electromagnetic waves that travel along the interface of a metal and a dielectric.  They are bound to the interface due to the mixing of the electromagnetic field oscillation with the oscillation of free electrons in the metal\cite{maier:plasmonics}.  Their unique properties draw interest from a variety of fields\cite{stockman:39}, such as nanophotonics\cite{ozbay:189}, solar cells\cite{atwater:205}, biological imaging and sensing\cite{homola:3,chung:10907}.  The fundamental property of plasmonic excitations is the scaling of the plasma resonance frequency  with the square root of the electron density: $\omega_p\propto \sqrt{n}$.  To observe plasma resonances at Terahertz (THz) frequencies, the electron density typically needs to be lowered down to the $n\sim 10^{15}-10^{16}$ cm$^{-3}$ range, which is achieved in undoped or lightly doped semiconductors.  For example, THz-frequency SPPs in doped Si were found responsible for extraordinary transmission of subwavelength hole arrays\cite{gomezrivas:201306,azad:141102}.  Alternatively, THz-frequency SPPs can exist on microscopically structured or corrugated metal surfaces\cite{williams:175,zhu:6216,fernandez:233104,maier:176805,ng:1059}.

Indium antimonide (InSb) is a semiconductor well suited for the study of THz SPPs.  It combines a temperature-tunable electron density with very high electron mobility, which is another requirement for the existence of well-defined SPPs.  Because of the low bandgap of InSb ($E_g=170$ meV), its intrinsic electron density at room temperature is about $10^{16}$ cm$^{-3}$, which corresponds to the bulk plasma frequency ($\omega_p^2=ne^2/\epsilon_0\epsilon_{\infty}m^*$) of $\omega_p/2\pi\sim1.8$ THz.  Propagating and localized surface plasmons on InSb surfaces and microstructures have been studied actively in recent years\cite{berg:55,gomezrivas:847,isaac:113411,zhu:3129,hanham:226,jung:1007,deng:128}, and the potential for THz sensing and spectroscopy has been pointed out\cite{isaac:241115}. 

\begin{figure}[ht!]
\centering\includegraphics[width=7cm]{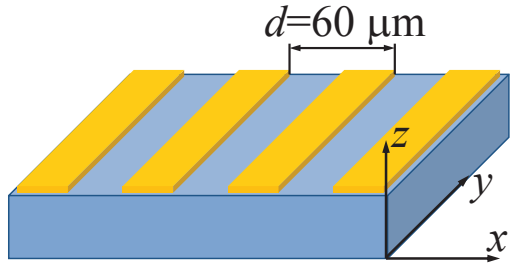}
\caption{\label{fig:geom}The gold grating and InSb layer structure is taken to stretch indefinitely in the $x$ and $y$ directions.  The THz wave impinges straight down along the negative $z$ direction and is polarized along $x$.}
\end{figure}

In this work, we show that a microscopically thin layer of InSb ($2-8\mc$ thick) combined with a metallic surface grating, Fig.~\ref{fig:geom}, can serve as a sensitive THz SPP sensor.  The micrometer-scale thickness allows the structure to operate in normal-incidence transmission mode, while the surface grating couples the incident THz radiation to the standing SPP waves at the InSb interface.  This transmission-mode operation sets this structure apart from the recent THz sensing schemes based on wave propagation along the interface\cite{isaac:241115,ng:1059} or in a waveguide\cite{nagel:s601}.  Our computational modeling of the THz optical properties of the grating/InSb structure shows the presence of two strong SPP resonances in the transmission spectrum.  This THz plasmonic response is highly sensitive to the dielectric environment at the InSb interface and propose a sensing modality that takes advantage of the inherent tunability of InSb SPP properties by temperature and/or doping.

Figure~\ref{fig:geom} shows the basic plasmonic structure consisting of a gold grating on the surface of a 5$\mc$ thick InSb sheet.  We take the grating with a period $d=60\mc$ to be periodic and continue indefinitely in the $x$ direction, and we take the whole structure to be infinitely long in the $y$ direction.  The THz wave is incident straight down along the negative $z$ direction.  The gold grating strips are 200 nm thick and  $30\mc$ wide with $30\mc$ gaps between them.  These dimensions place the grating comfortably within the range of conventional photolithographic fabrication methods.  As the grating couples the incident THz wave to the SPP modes at InSb interfaces, its periodicity also determines the fundamental wavevector $\beta_0$ of the standing SPP modes: $\beta_0=2\pi/d$.  Thus, we can model the dispersion of the observed SPP modes by varying the grating period $d$; we keep the strip widths and the strip gaps equal to each other in all simulations. 

\begin{figure}[ht!]
\centering\includegraphics[width=10cm]{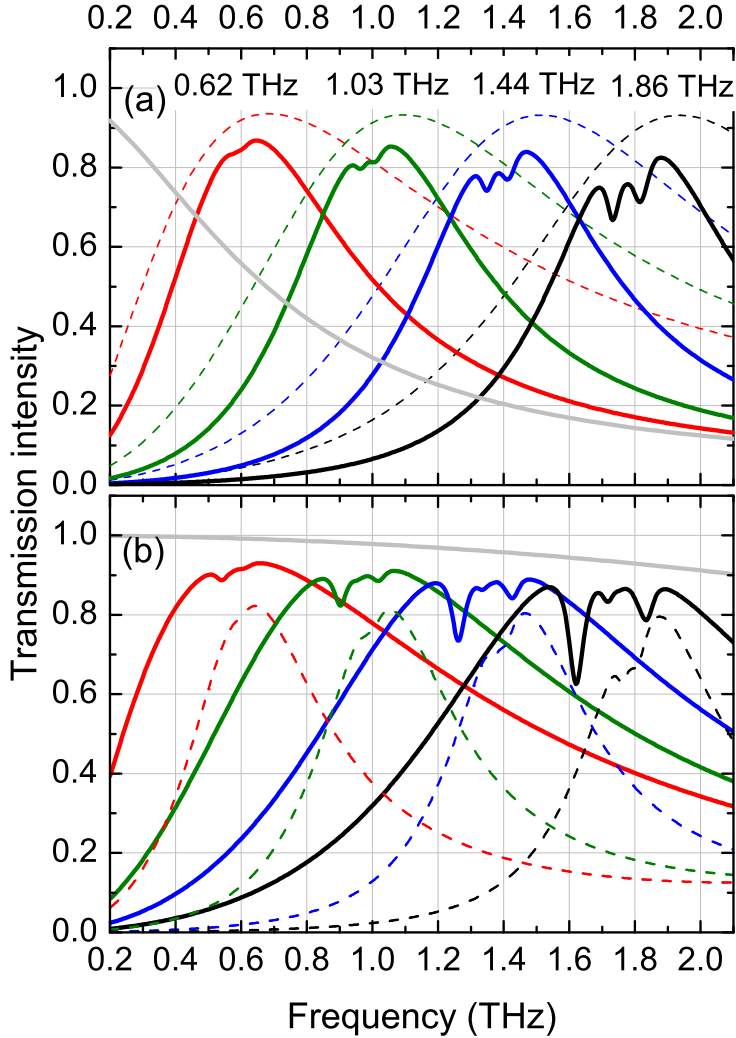}
\caption{\label{fig:trans1}(a) Transmission of the $5\mc$-thick InSb layer with (thick solid lines) and without (dashed lines) the gold grating as a function of the bulk plasma frequency, which is indicated by the labels above each spectrum.  The light-gray line shows the transmission of the structure with $\omega_p=0$. (b) Transmission of the $2\mc$-thick (thick solid lines) and $8\mc$-thick (dashed lines) InSb layer with gold grating as a function of the bulk plasma frequency. The light-gray line shows the transmission of a free-standing grating. The grating period is $d=60\mc$ for all curves in (a) and (b).  }
\end{figure}

\section{Results and discussion}
\subsection{Plasmonic response of the gold grating/InSb structure}
THz optical properties of the grating/InSb structure were modeled using commercial software packages COMSOL Multiphysics and CST Microwave Studio.  Both packages returned consistent simulation results.  The simulation model was built in 2D to maximize the computation speed, by taking the structure to be infinite and translationally invariant in the $y$ direction, Fig.~\ref{fig:geom}.  The simulations were performed by the COMSOL RF module's stationary solver in the frequency domain.  We achieved identical results with either the periodic boundary conditions or the perfect electric conductor (PEC) boundary conditions.  The material above and below the structure is air (or vacuum).  The incident THz electric field is polarized in the $x$ direction.  The complex dielectric permittivity of both InSb and gold is approximated by the Drude model, Eq. (\ref{eq:1}).  The gold Drude parameters are $\epsilon_\infty=9.1$, $\omega_p/2\pi=728$ THz, and $\gamma=1.1\times10^{14}$ rad/s.

Figure~\ref{fig:trans1}(a) shows the transmission of the $5\mc$ InSb layer with and without the gold grating.  The overall transmission exhibit a broad peak shape in both cases, with the peak position roughly following the bulk plasma frequency.  Such band-pass structure of the overall transmission in this frequency range results from the Fabry-Perot effect between the top and bottom surfaces of the InSb layer and from the Drude response of conduction electrons.  The low transmission on the zero-frequency side of the peak is due to the reflection and absorption of the THz wave according to the Drude model.  The drop-off in transmission on the high-frequency side of the peak results from the onset of the first Fabry-Perot minimum.  The peak position is roughly determined by the plasma frequency, as it sets the width of the low-transmission frequency window of the electron Drude response.  The transmission peaks are narrower for the thicker InSb layers, because the low-transmission window on the zero-frequency side is sharper and deeper, while the separation between Fabry-Perot fringes is smaller, thus resulting in the faster transmission drop-off on the high-frequency side.

In addition to the broad peak, two clear resonance dips appear in the transmission spectrum in Fig.~\ref{fig:trans1}(a), as compared to the bare $5\mc$-thick InSb layer without the gold grating.  The position of the resonant dips strongly depends on the bulk plasma frequency $\omega_p$, which gives the first indication that these resonances are associated with standing SPPs at the InSb/air interfaces.  Figure~\ref{fig:trans1}(b) shows the dependence of the observed SPP resonances on the InSb thickness.  As the thickness gets smaller, we observe that: (i) The resonances get stronger.  (ii) The separation between the resonances increases.  (iii) The lower-frequency resonance shifts to a lower frequency.  As we show below, this phenomenology is consistent with the theoretical properties of the SPPs in the model of air/InSb/air trilayer without the metal grating, while also exhibiting significant differences with this simplified model.

In the air/InSb/air trilayer theory, the propagating SPP modes exist at each of the two InSb/air interfaces.  For sufficiently thick InSb, these SPPs are completely independent and identical.  They are bound to the interface, with the decay lengths $z_1$ and $z_2$ in InSb and air perpendicular to the interface.  The inverse quantities $k_i=1/z_i$ ($i=1,2$) are the imaginary wavevectors in the directions perpendicular to the interface.  As the InSb thickness gets small, the SPPs at the two interfaces interact and split into even and odd modes based on the parity of the $E_x(z)$ field.  In the interaction regime, the SPP dispersion relations are given by\cite{maier:plasmonics}
\begin{equation}
k_i^2=\beta^2-k_0^2\epsilon_i\\
\label{eq:beta}
\end{equation}
and by
\begin{eqnarray}
\label{eq:odd}
\tanh k_1a = -\frac{k_2\epsilon_1}{k_1\epsilon_2},\\
\label{eq:even}
\tanh k_1a = -\frac{k_1\epsilon_2}{k_2\epsilon_1},
\end{eqnarray}
where $k_0=\omega/c$, $\beta$ is the SPP wavevector along the interface, $\epsilon_1$ and $\epsilon_2$ are the dielectric permittivities of InSb and air, and $a$ is the thickness of the InSb layer.  The complex dielectric permittivity of InSb is approximated by the Drude model
\begin{equation}
\epsilon(\omega)=\epsilon_{\infty}(1-\frac{\omega_p^2}{\omega^2+i\omega\gamma}),
\label{eq:1}
\end{equation}
where is the bulk plasma frequency, $\gamma$ is the electron scattering rate, and $\epsilon_{\infty}$ is the background dielectric constant.  We use the low-temperature InSb scattering rate of $\gamma=0.3\times 10^{12}$ rad/s, which was confirmed to be relatively independent of plasma frequency in the range of interest.  InSb background dielectric constant is $\epsilon_{\infty}=15.6$.

\begin{figure}[ht!]
\centering\includegraphics[width=10cm]{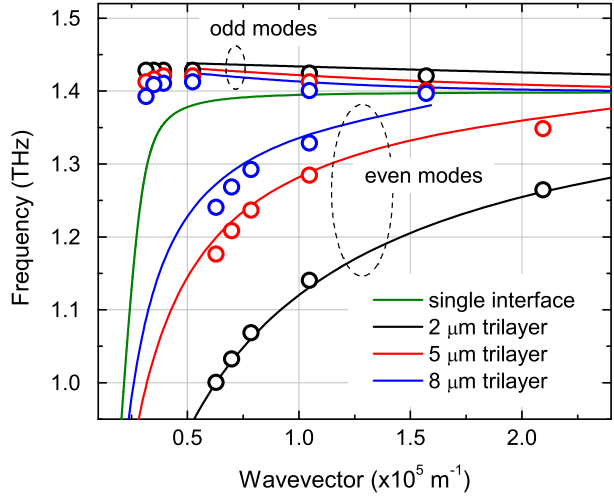}
\caption{\label{fig:disp}Open circles - computational frequency vs. wavevector $\omega(\beta)$ dispersion of the resonances on the grating/InSb structure for different InSb thicknesses.  Solid lines - thickness-dependent theoretical dispersion of SPP modes in the air/InSb/air trilayer structure.  The single interface line shows the theoretical SPP dispersion on a single InSb/air interface without a grating, where both air and InSb assume infinite thickness.  The bulk plasma frequency of InSb is $\omega_p=1.44$ THz.}
\end{figure}

Equations (\ref{eq:odd}) and (\ref{eq:even}) describe the SPP modes of odd and even parity, respectively\cite{maier:plasmonics}.  Figure~\ref{fig:disp} shows the theoretical dispersion curves calculated from Eqs. (\ref{eq:beta}-\ref{eq:even}) for the plasma frequency $\omega_p=1.44$ THz and several thicknesses of the InSb layer.  The dispersion of even modes was calculated by solving the system of Eqs. (\ref{eq:beta}) and (\ref{eq:even}).  The odd mode dispersion was calculated by setting $\gamma=0$ in InSb and assuming that the wavevector $\beta\gg\omega_p/c$.  In this case, we can take $k_1\simeq k_2 \simeq \beta$ and the odd mode dispersion for large wavevectors can be calculated from
\begin{equation}
\omega=\frac{\omega_p}{\left[ 1+(\epsilon_2/\epsilon_\infty)\tanh(a\beta)\right]^{1/2} },
\label{eq:oddlargeb}
\end{equation}
where $a$ is the thickness of the InSb slab.  In Fig.~\ref{fig:disp}, we used the effective thickness $a$ equal to twice the actual InSb thickness.  This is necessitated by the spatial distribution of the observed electric field $E_z$ in the higher SPP resonance, as we will discuss below.

\begin{figure}[ht!]
\centering\includegraphics[width=10cm]{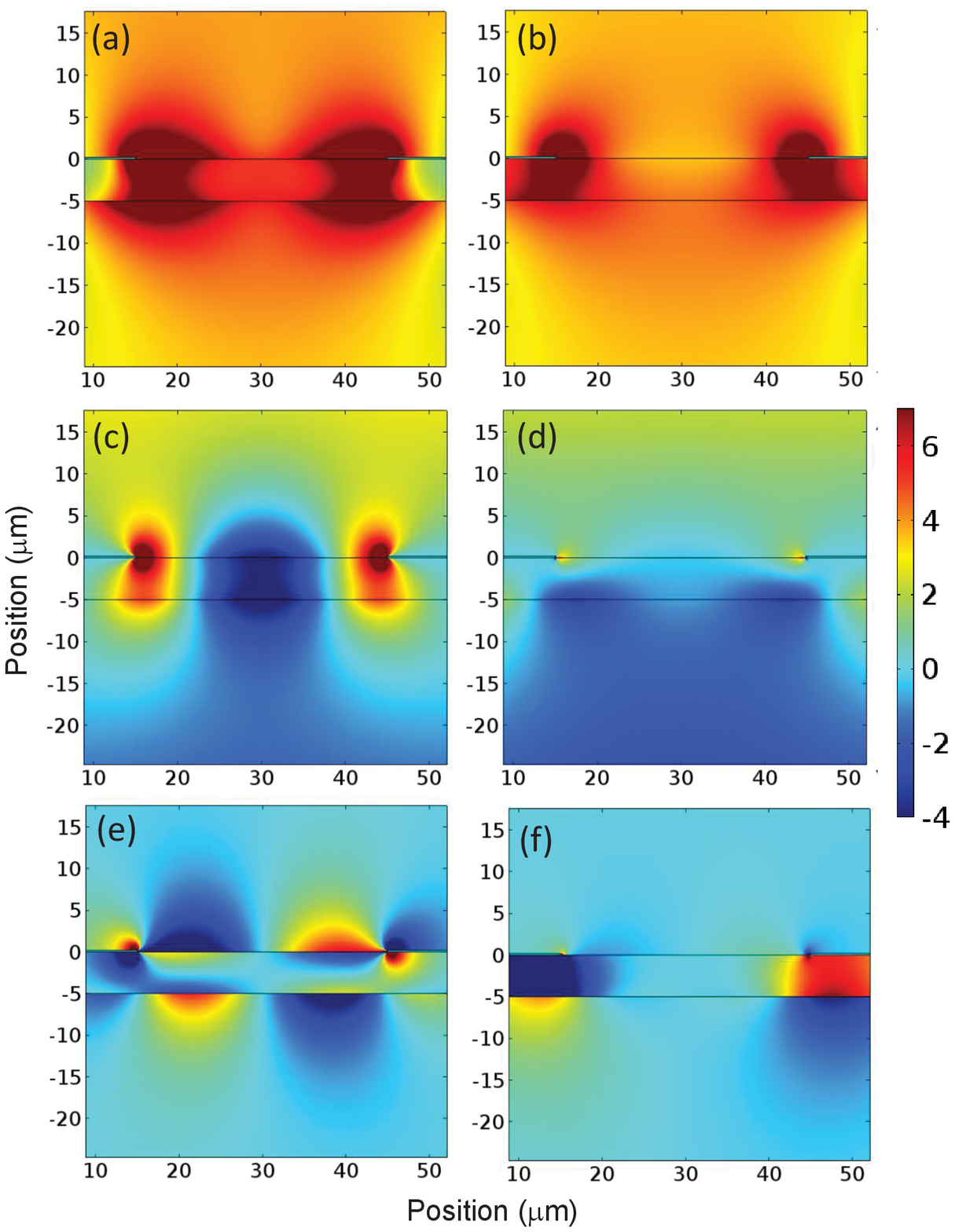}
\caption{\label{fig:field}(a),(b) Spatial distribution of the electric field amplitude for the lower (1.36 THz) and higher (1.42 THz) SPP modes in the $5\mc-$thick InSb with grating period $d=60\mc$.  The bulk plasma frequency is $\omega_p=1.44$ THz.  The color bar indicates the relative amplitude scale using arbitrary units.  The InSb layer fills the vertical ($-5\mc$,$0\mc$) interval.  The gold strips cover the horizontal ($0\mc$,$15\mc$) and ($45\mc$,$60\mc$) intervals at the top surface of InSb.  (c),(d) The spatial distribution of the $E_x$ component of the electric field for the lower and higher SPP modes.  (e),(f) The spatial distribution of the $E_z$ component of the electric field for the lower and higher SPP modes.}
\end{figure}

We compare the theoretical dispersion curves with the resonance frequencies found in our numerical simulation, where the SPP wavevector is determined by the grating period $d$.  By computing the transmission of our structure with different grating periods $d$, we can determine the dispersion $\omega(\beta)$ of the observed SPP resonances.  The examination of the electric field distribution at the two resonance frequencies shows that the field's spatial dependence can be represented as a Fourier series with the fundamental wavevector $\beta_0=2\pi/d$.  The largest-amplitude term in the series for the lower-frequency SPP resonance is the second harmonic $2\beta_0$, which we use to plot the computationally determined dispersion in Fig.~\ref{fig:disp}.  The largest-amplitude term in the series for the high-frequency resonance is the fundamental wavevector $\beta_0$, and it is used to plot the computational dispersion of this resonance.  Figure~\ref{fig:disp} shows a good agreement between the theoretical dispersion (solid lines) and the computationally observed one (symbols), indicating that the trilayer theory correctly describes the main phenomenology of the observed SPP resonances.  Therefore, we assign the two strong absorption resonances in the spectra of Fig.~\ref{fig:trans1} as the even and odd SPP modes that exist on the surfaces of the air/InSb/air trilayer.

The separation between the even and odd SPP modes decreases with the bulk plasma frequency, as is evident in Fig.~\ref{fig:trans1}.  As the plasma frequency is decreased, the two modes merge into one resonant line.  At $\omega_p=0.62$ THz, only a single line appears in the spectra because the even and odd modes have merged into one.  When the separation between the even and odd modes is large and the resonances are strong, a week resonant dip appears in the spectra in-between the main even and odd SPP modes, Fig.~\ref{fig:trans1}. It can be interpreted as an overtone of the lower even SPP mode. 

Figures~\ref{fig:field}(a,b) show the electric field amplitude distribution at both SPP frequencies in the structure with $\omega_p=1.44$ THz, while Figs.~\ref{fig:field}(c,d) and Figs.~\ref{fig:field}(e,f) show the corresponding spatial distributions of the $x$-component ($E_x$) and the $z$-component ($E_z$) of the electric field.  In the lower-frequency (1.36 THz) SPP mode, the electric field is concentrated in the gap between the gold strips of the grating. The $x$-component shows the same sign, Fig.~\ref{fig:field}(c), and the $z$-component shows the opposite sign, Fig.~\ref{fig:field}(e), at the upper and lower InSb interface.  Such behavior corresponds to the even mode in the simple  trilayer theory and further justifies our assignment of the lower SPP resonance as the even mode.  The electric field of the higher-frequency (1.42 THz) SPP mode is the highest under the gold strips inside InSb, which is most evident for the $z$-component $E_z$ in Fig.~\ref{fig:field}(f).  The spatial distribution of the field $E_z$ under the gold strips is consistent with the field of the odd mode in an air/InSb/air structure that is twice as thick.  That is, if we "reflect" the field $E_z$ underneath the gold strip about the gold/InSb interface and into the upper half-space, we will obtain the correct field distribution of the even mode of the air/InSb/air trilayer with twice the thickness.  For this reason, we used the effective thickness of 2$a$ in Eq. (\ref{eq:oddlargeb}) to determine the theoretical odd mode dispersion.

\begin{figure}[ht!]
\centering\includegraphics[width=10cm]{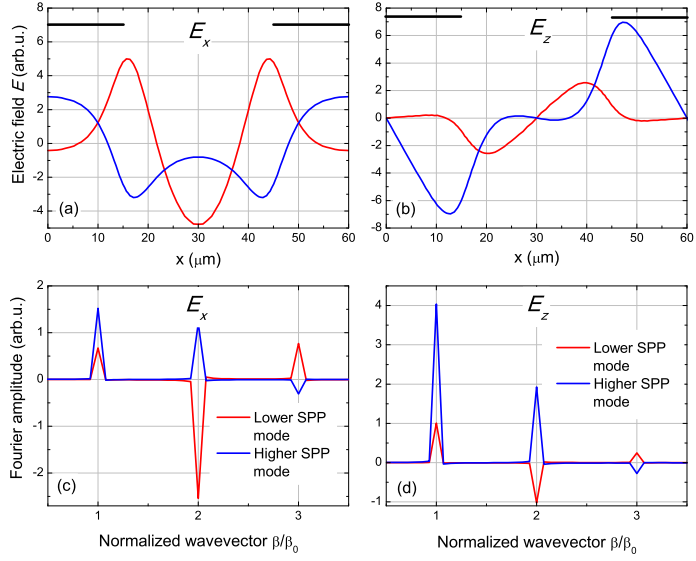}
\caption{\label{fig:cut} (a),(b) The electric field components $E_x(x)$ and $E_z(x)$ along the lower InSb surface in the $5\mc$-thick InSb structure at the lower (1.36 THz) and higher (1.42 THz) SPP modes.  The grating period is $d=60\mc$ ($\beta_0=2\pi/d$) and bulk plasma frequency is $\omega_p=1.44$ THz.  The horizontal black lines indicate the position of the grating gold strips on the upper surface of the InSb wafer. (c),(d) Amplitudes of the Fourier series expansion for the electric fields $E_x(x)$ and $E_z(x)$ shown in (a) and (b).  In (c), only the cosine terms are shown, and the sine terms are all zero.  In (d), only the sine terms are shown, and the cosine terms are all zero.}
\end{figure}

To gain further insight into the nature of the SPP modes in our structure, we plot the variation of the fields $E_x$ and $E_z$ along the $x$ direction in Figs.~\ref{fig:cut}(a,b), where $\omega_p=1.44$ THz.  Since the grating is periodic in this direction, both fields can be written as Fourier series in terms of the fundamental wavevector $\beta_0=2\pi/d$, where $d$ is the grating period.  At the edge of the unit cell $E_z=0$, therefore $E_z(x)$ only contains sine terms in the Fourier series. Similarly, $E_x(x)$ reaches a local maximum (or minimum) value at the edge, so it only contains cosine terms.  Figures~\ref{fig:cut}(c,d) show the relative amplitudes of the main terms in Fourier series expansion of $E_x(x)$ and $E_z(x)$.  The largest-amplitude term in the lower SPP mode is the second harmonic with the wavevector $2\beta_0$, with terms up to the third order being significant in both cases.  We used the second harmonic wavevector to plot the even mode dispersion in Fig.~\ref{fig:disp}.  The opposite sign of the first and second harmonic amplitudes of the lower SPP mode (red line in Fig.~\ref{fig:cut}(c)) indicates that the first and second harmonic cosine terms add constructively in the gap between the gold strips, while they add destructively underneath the strips.  Similarly in Fig.~\ref{fig:cut}(d), the first and second harmonic sine terms of the lower SPP mode (red line) add constructively in the gap and destructively underneath the strips.  At off-resonance frequencies, the electric field $E_z(x)$ becomes negligible.  The simulation shows that the absorption resonances appear in the transmission spectrum only when the incident field excites a high amplitude $E_z(x)$ component, which is another clear indication of the SPP origin of the absorption. 

\subsection{Potential for sensor applications}
\begin{figure}[ht!]
\centering\includegraphics[width=10cm]{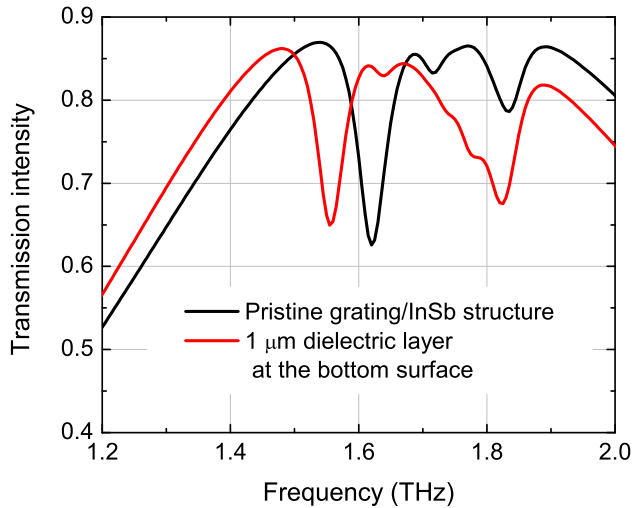}
\caption{\label{fig:6}Transmission of a $2\mc$ grating/InSb structure with (red) and without (black) a $1\mc$ dielectric layer at the bottom InSb surface.  The THz refractive index of the dielectric layer is $n=2$.  InSb bulk plasma frequency is $\omega_p=1.86$ THz.  The grating period is $d=60\mc$.}
\end{figure}

We explored the potential of our grating/InSb structure for sensing applications.  Typical sensors are based on the modulation of the resonant plasmonic response by the dielectric environment\cite{chung:10907}.  We computed the THz transmission of the $2\mc$ grating/InSb structure in the presence of a thin $1\mc$ dielectric layer with THz refractive index $n=2$ at the lower InSb/air interface, and found a sharp difference in the resonant transmission response, Fig.~\ref{fig:6}.  The structure with the dielectric layer exhibits the same two SPP resonances.  However, the lower SPP resonance exhibits a clear and significant shift to a lower frequency.  The higher SPP resonance does not change its frequency, but becomes dramatically stronger, with the area under the resonance peak increasing by a factor of 4.  The on-resonance transmitted THz intensity is about 15\% lower for the structure with the dielectric layer.  Our results demonstrate that the grating/InSb plasmonic structure can act as a sensitive probe of THz-frequency dielectric environment with sensitivity down to sub-$\mc$ dielectric layers.

The plasmonic sensor sensitivity is usually quantified as $\eta=\Delta\lambda/\Delta n$ (in nm per refractive index unit, nm/RIU) or as $\eta=\Delta f/\Delta n$ (in Hz/RIU).  For the lower SPP resonance in our structure, we get $\eta \approx7200$ nm/RIU (or $\eta\approx0.06$ THz/RIU).  This sensitivity is lower than that of the sensors based on guided surface plasmon waves, such as those propagating on spoof THz plasmon surfaces\cite{ng:2195} (0.5 THz/RIU).  The higher sensitivity of the sensors based on the propagating SPPs\cite{ng:2195,ng:1059} results from an increased interaction length of the SPP wave and the analyte along the guiding structure.  That sensitivity is not matched in our sensor, as it is based on normal-incidence transmission geometry.  However, the simpler optical architecture that does not require the coupling of free-space radiation to the guided SPP modes and the fabrication that relies on conventional photolithography may make our sensor competitive in many application areas.  

While the intimate link between the dielectric environment and the surface plasmon resonance enables the delicate sensitivity of plasmonic sensors, their selectivity often needs to be engineered by chemically functionalizing the plasmonic structures to be receptive only to a specific analyte\cite{homola:3,chung:10907,stockman:39}.  By contrast, THz plasmonic sensors may allow us to implement both sensitivity and selectivity using the inherent properties of THz SPPs, because many molecular materials exhibit very specific spectroscopic fingerprints at THz frequencies, mostly of vibrational origin\cite{melinger:a79,ng:1059,brown:061908,walther:261107}.  Molecular detection and marker-free biosensing schemes based on such THz fingerprints have been proposed\cite{melinger:a79,ng:1059,walther:261107,nagel:s601}.

\begin{figure}[ht!]
\centering\includegraphics[width=10cm]{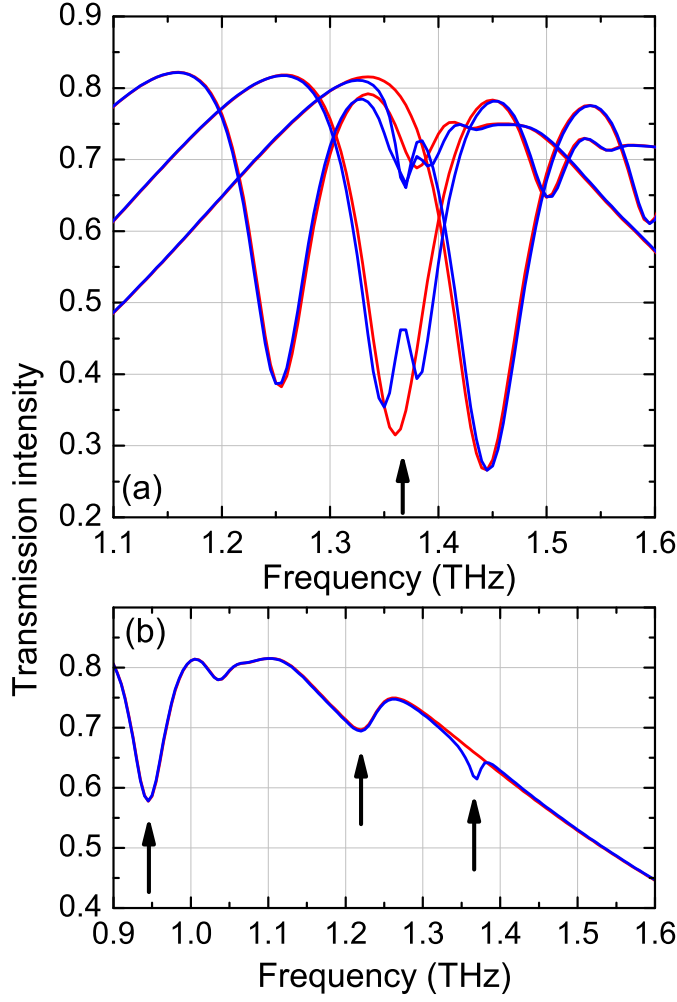}
\caption{\label{fig:7}(a) Interaction of the lower-frequency SPP mode with the lactose vibrational resonance at 1.37 THz (the arrow).  The lower SPP mode is shown for the bulk plasma frequencies of $\omega_p=1.65$ THz, $1.79$ THz, and $1.91$ THz.  The red lines show the transmission without the vibrational resonance (resonance strength set to zero).  The blue lines show the transmission with the vibrational resonance.  The InSb thickness is $2\mc$ and the grating period is $d=120\mc$.  (b) Transmission of the grating/InSb structure with the $1\mc$ lactose layer with (blue line) and without (red line) the vibrational resonance when the SPP modes are detuned far away from 1.37 THz.  The arrows indicate the lower and higher SPP modes and the lactose resonance at 1.37 THz.}
\end{figure}

We examine the potential of our grating/InSb structure for THz molecular selectivity.  We use lactose ($\alpha$-lactose monohydrate) as a model molecular material, as it exhibits a strong vibrational resonance at 1.37 THz\cite{ng:1059,brown:061908,walther:261107}.  Figure~\ref{fig:7} shows the computed transmission of the grating/InSb structure with a $1\mc$ thick lactose layer on the bottom surface. We model the lactose optical properties in the Lorentz model using the background dielectric constant $\epsilon_{\infty}=4$ (background refractive index $n=2$) and a realistic Lorentz oscillator strength\cite{brown:061908}.  We use the Lorentz model damping rate of $0.1\times10^{12}$ rad/s, which is exhibited by thin polycrystalline molecular films at low temperature\cite{melinger:a79}.  For comparison, Fig.~\ref{fig:7} also shows the transmission of the structure when the Lorentz oscillator strength is set to zero.  When the SPP resonance is tuned away from 1.37 THz vibrational frequency, the vibrational resonance results in a 6\% drop in transmitted intensity, Fig.~\ref{fig:7}(b).  When the lower SPP frequency is tuned in resonance with the vibrational frequency, the SPP absorption line splits into two absorption lines, with a significant increase in transmitted intensity, Fig.~\ref{fig:7}(a).  On resonance, the transmitted intensity increases by almost 50\%, which is significantly higher than the 6\% difference found away from the SPP resonance. The splitting of the lower SPP resonance suggests a coupling between the SPP and vibrational modes of the analyte (lactose).  This coupling distinguishes our sensor from other propagating or localized surface plasmon sensors and suggests a new sensing modality, in which the SPP resonance is tuned to sample the frequencies in the vicinity of 1.37 THz.  By measuring the shape and strength of the lower SPP absorption as the SPP frequency changes, we can distinguish the lactose layer at the bottom surface from other materials that possess a similar background dielectric constant but do not exhibit the sharp resonant absorption at 1.37 THz, Fig.~\ref{fig:7}(a).  In a similar fashion, this sensing modality can be employed for marker-free detection of other molecules with specific THz fingerprints.

\section{Conclusions}
We have explored the THz-frequency optical properties of a micrometer-thin InSb slab with a gold grating and found a strong resonant response consisting of two SPP modes.  Their dispersion and dependence on InSb thickness are well described by the theory of SPPs propagating in the simple trilayer structure without the gold grating.  The optical properties of the grating/InSb structure are highly sensitive to the presence of an analyte at the lower InSb interface, which allows potential applications as a THz plasmonic sensor.  We have also found a coupling between the lower SPP mode and the vibrational mode of lactose and proposed a sensing modality that provides marker-free selectivity based on the THz vibrational spectral signatures of the analyte.

\section*{Funding}
Louisiana Board of Regents, the Board of Regents Support Fund (LEQSF(2012-15)-RD-A-23); Alfred P. Sloan Foundation (BR2013-123); KRISS (GP2016-034).

%%%%%%%%%%%%%%%%%%%%%%% References %%%%%%%%%%%%%%%%%%%%%%%%%
%\bibliography{insbgrating} 

\begin{thebibliography}{26}
\expandafter\ifx\csname natexlab\endcsname\relax\def\natexlab#1{#1}\fi
\expandafter\ifx\csname bibnamefont\endcsname\relax
  \def\bibnamefont#1{#1}\fi
\expandafter\ifx\csname bibfnamefont\endcsname\relax
  \def\bibfnamefont#1{#1}\fi
\expandafter\ifx\csname citenamefont\endcsname\relax
  \def\citenamefont#1{#1}\fi
\expandafter\ifx\csname url\endcsname\relax
  \def\url#1{\texttt{#1}}\fi
\expandafter\ifx\csname urlprefix\endcsname\relax\def\urlprefix{URL }\fi
\providecommand{\bibinfo}[2]{#2}
\providecommand{\eprint}[2][]{\url{#2}}

\bibitem[{\citenamefont{Maier}(2007)}]{maier:plasmonics}
\bibinfo{author}{\bibfnamefont{S.~A.} \bibnamefont{Maier}},
  \emph{\bibinfo{title}{Plasmonics: Fundamentals and Applications}}
  (\bibinfo{publisher}{Springer}, \bibinfo{year}{2007}), ISBN
  \bibinfo{isbn}{0-387-33150-6}.

\bibitem[{\citenamefont{Stockman}(2011)}]{stockman:39}
\bibinfo{author}{\bibfnamefont{M.~I.} \bibnamefont{Stockman}},
  \bibinfo{journal}{Physics Today} \textbf{\bibinfo{volume}{64}}
  (\bibinfo{year}{2011}).

\bibitem[{\citenamefont{Ozbay}(2006)}]{ozbay:189}
\bibinfo{author}{\bibfnamefont{E.}~\bibnamefont{Ozbay}},
  \bibinfo{journal}{Science} \textbf{\bibinfo{volume}{311}},
  \bibinfo{pages}{189} (\bibinfo{year}{2006}), ISSN \bibinfo{issn}{0036-8075},
  \urlprefix\url{http://science.sciencemag.org/content/311/5758/189}.

\bibitem[{\citenamefont{Atwater and Polman}(2010)}]{atwater:205}
\bibinfo{author}{\bibfnamefont{H.}~\bibnamefont{Atwater}} \bibnamefont{and}
  \bibinfo{author}{\bibfnamefont{A.}~\bibnamefont{Polman}},
  \bibinfo{journal}{Nature Mater.} \textbf{\bibinfo{volume}{9}},
  \bibinfo{pages}{205} (\bibinfo{year}{2010}),
  \urlprefix\url{http://www.nature.com/nmat/journal/v9/n3/full/nmat2629.html}.

\bibitem[{\citenamefont{Homola et~al.}(1999)\citenamefont{Homola, Yee, and
  Gauglitz}}]{homola:3}
\bibinfo{author}{\bibfnamefont{J.}~\bibnamefont{Homola}},
  \bibinfo{author}{\bibfnamefont{S.~S.} \bibnamefont{Yee}}, \bibnamefont{and}
  \bibinfo{author}{\bibfnamefont{G.}~\bibnamefont{Gauglitz}},
  \bibinfo{journal}{Sensors and Actuators B: Chemical}
  \textbf{\bibinfo{volume}{54}}, \bibinfo{pages}{3 } (\bibinfo{year}{1999}),
  ISSN \bibinfo{issn}{0925-4005},
  \urlprefix\url{http://www.sciencedirect.com/science/article/pii/S0925400598003219}.

\bibitem[{\citenamefont{Chung et~al.}(2011)\citenamefont{Chung, Lee, Song, H.,
  and Lee}}]{chung:10907}
\bibinfo{author}{\bibfnamefont{T.}~\bibnamefont{Chung}},
  \bibinfo{author}{\bibfnamefont{S.}~\bibnamefont{Lee}},
  \bibinfo{author}{\bibfnamefont{E.}~\bibnamefont{Song}},
  \bibinfo{author}{\bibfnamefont{C.}~\bibnamefont{H.}}, \bibnamefont{and}
  \bibinfo{author}{\bibfnamefont{B.}~\bibnamefont{Lee}},
  \bibinfo{journal}{Sensors} \textbf{\bibinfo{volume}{11}},
  \bibinfo{pages}{10907} (\bibinfo{year}{2011}),
  \urlprefix\url{http://www.mdpi.com/1424-8220/11/11/10907}.

\bibitem[{\citenamefont{G\'omez~Rivas et~al.}(2003)\citenamefont{G\'omez~Rivas,
  Schotsch, Haring~Bolivar, and Kurz}}]{gomezrivas:201306}
\bibinfo{author}{\bibfnamefont{J.}~\bibnamefont{G\'omez~Rivas}},
  \bibinfo{author}{\bibfnamefont{C.}~\bibnamefont{Schotsch}},
  \bibinfo{author}{\bibfnamefont{P.}~\bibnamefont{Haring~Bolivar}},
  \bibnamefont{and} \bibinfo{author}{\bibfnamefont{H.}~\bibnamefont{Kurz}},
  \bibinfo{journal}{Phys. Rev. B} \textbf{\bibinfo{volume}{68}},
  \bibinfo{pages}{201306} (\bibinfo{year}{2003}),
  \urlprefix\url{http://link.aps.org/doi/10.1103/PhysRevB.68.201306}.

\bibitem[{\citenamefont{Azad et~al.}(2005)\citenamefont{Azad, Zhao, and
  Zhang}}]{azad:141102}
\bibinfo{author}{\bibfnamefont{A.~K.} \bibnamefont{Azad}},
  \bibinfo{author}{\bibfnamefont{Y.}~\bibnamefont{Zhao}}, \bibnamefont{and}
  \bibinfo{author}{\bibfnamefont{W.}~\bibnamefont{Zhang}},
  \bibinfo{journal}{Applied Physics Letters} \textbf{\bibinfo{volume}{86}},
  \bibinfo{eid}{141102} (\bibinfo{year}{2005}),
  \urlprefix\url{http://scitation.aip.org/content/aip/journal/apl/86/14/10.1063/1.1897842}.

\bibitem[{\citenamefont{Williams et~al.}(2008)\citenamefont{Williams, Andrews,
  Maier, Fernandez-Dominguez, Martin-Moreno, and Garcia-Vidal}}]{williams:175}
\bibinfo{author}{\bibfnamefont{C.}~\bibnamefont{Williams}},
  \bibinfo{author}{\bibfnamefont{S.}~\bibnamefont{Andrews}},
  \bibinfo{author}{\bibfnamefont{S.}~\bibnamefont{Maier}},
  \bibinfo{author}{\bibfnamefont{A.}~\bibnamefont{Fernandez-Dominguez}},
  \bibinfo{author}{\bibfnamefont{L.}~\bibnamefont{Martin-Moreno}},
  \bibnamefont{and}
  \bibinfo{author}{\bibfnamefont{F.}~\bibnamefont{Garcia-Vidal}},
  \bibinfo{journal}{Nat. Photon.} \textbf{\bibinfo{volume}{2}},
  \bibinfo{pages}{175} (\bibinfo{year}{2008}),
  \urlprefix\url{http://www.nature.com/nphoton/journal/v2/n3/abs/nphoton.2007.301.html}.

\bibitem[{\citenamefont{Zhu et~al.}(2008)\citenamefont{Zhu, Agrawal, and
  Nahata}}]{zhu:6216}
\bibinfo{author}{\bibfnamefont{W.}~\bibnamefont{Zhu}},
  \bibinfo{author}{\bibfnamefont{A.}~\bibnamefont{Agrawal}}, \bibnamefont{and}
  \bibinfo{author}{\bibfnamefont{A.}~\bibnamefont{Nahata}},
  \bibinfo{journal}{Opt. Express} \textbf{\bibinfo{volume}{16}},
  \bibinfo{pages}{6216} (\bibinfo{year}{2008}),
  \urlprefix\url{http://www.opticsexpress.org/abstract.cfm?URI=oe-16-9-6216}.

\bibitem[{\citenamefont{Fern\'andez-Dom\'{\i}nguez
  et~al.}(2009)\citenamefont{Fern\'andez-Dom\'{\i}nguez, Moreno,
  Mart\'{\i}n-Moreno, and Garc\'{\i}a-Vidal}}]{fernandez:233104}
\bibinfo{author}{\bibfnamefont{A.~I.}
  \bibnamefont{Fern\'andez-Dom\'{\i}nguez}},
  \bibinfo{author}{\bibfnamefont{E.}~\bibnamefont{Moreno}},
  \bibinfo{author}{\bibfnamefont{L.}~\bibnamefont{Mart\'{\i}n-Moreno}},
  \bibnamefont{and} \bibinfo{author}{\bibfnamefont{F.~J.}
  \bibnamefont{Garc\'{\i}a-Vidal}}, \bibinfo{journal}{Phys. Rev. B}
  \textbf{\bibinfo{volume}{79}}, \bibinfo{pages}{233104}
  (\bibinfo{year}{2009}),
  \urlprefix\url{http://link.aps.org/doi/10.1103/PhysRevB.79.233104}.

\bibitem[{\citenamefont{Maier et~al.}(2006)\citenamefont{Maier, Andrews,
  Mart\'{\i}n-Moreno, and Garc\'{\i}a-Vidal}}]{maier:176805}
\bibinfo{author}{\bibfnamefont{S.~A.} \bibnamefont{Maier}},
  \bibinfo{author}{\bibfnamefont{S.~R.} \bibnamefont{Andrews}},
  \bibinfo{author}{\bibfnamefont{L.}~\bibnamefont{Mart\'{\i}n-Moreno}},
  \bibnamefont{and} \bibinfo{author}{\bibfnamefont{F.~J.}
  \bibnamefont{Garc\'{\i}a-Vidal}}, \bibinfo{journal}{Phys. Rev. Lett.}
  \textbf{\bibinfo{volume}{97}}, \bibinfo{pages}{176805}
  (\bibinfo{year}{2006}),
  \urlprefix\url{http://link.aps.org/doi/10.1103/PhysRevLett.97.176805}.

\bibitem[{\citenamefont{Ng et~al.}(2014)\citenamefont{Ng, Hanham, Wu,
  FernÃ¡ndez-DomÃ­nguez, Klein, Liew, Breese, Hong, and Maier}}]{ng:1059}
\bibinfo{author}{\bibfnamefont{B.}~\bibnamefont{Ng}},
  \bibinfo{author}{\bibfnamefont{S.~M.} \bibnamefont{Hanham}},
  \bibinfo{author}{\bibfnamefont{J.}~\bibnamefont{Wu}},
  \bibinfo{author}{\bibfnamefont{A.~I.} \bibnamefont{FernÃ¡ndez-DomÃ­nguez}},
  \bibinfo{author}{\bibfnamefont{N.}~\bibnamefont{Klein}},
  \bibinfo{author}{\bibfnamefont{Y.~F.} \bibnamefont{Liew}},
  \bibinfo{author}{\bibfnamefont{M.~B.~H.} \bibnamefont{Breese}},
  \bibinfo{author}{\bibfnamefont{M.}~\bibnamefont{Hong}}, \bibnamefont{and}
  \bibinfo{author}{\bibfnamefont{S.~A.} \bibnamefont{Maier}},
  \bibinfo{journal}{ACS Photonics} \textbf{\bibinfo{volume}{1}},
  \bibinfo{pages}{1059} (\bibinfo{year}{2014}),
  \eprint{http://dx.doi.org/10.1021/ph500272n},
  \urlprefix\url{http://dx.doi.org/10.1021/ph500272n}.

\bibitem[{\citenamefont{van~den Berg and Borburgh}(1974)}]{berg:55}
\bibinfo{author}{\bibfnamefont{P.}~\bibnamefont{van~den Berg}}
  \bibnamefont{and} \bibinfo{author}{\bibfnamefont{J.}~\bibnamefont{Borburgh}},
  \bibinfo{journal}{Appl. Phys.} \textbf{\bibinfo{volume}{3}},
  \bibinfo{pages}{55} (\bibinfo{year}{1974}).

\bibitem[{\citenamefont{Rivas et~al.}(2005)\citenamefont{Rivas, Janke, Bolivar,
  and Kurz}}]{gomezrivas:847}
\bibinfo{author}{\bibfnamefont{J.~G.} \bibnamefont{Rivas}},
  \bibinfo{author}{\bibfnamefont{C.}~\bibnamefont{Janke}},
  \bibinfo{author}{\bibfnamefont{P.~H.} \bibnamefont{Bolivar}},
  \bibnamefont{and} \bibinfo{author}{\bibfnamefont{H.}~\bibnamefont{Kurz}},
  \bibinfo{journal}{Opt. Express} \textbf{\bibinfo{volume}{13}},
  \bibinfo{pages}{847} (\bibinfo{year}{2005}),
  \urlprefix\url{http://www.opticsexpress.org/abstract.cfm?URI=oe-13-3-847}.

\bibitem[{\citenamefont{Isaac et~al.}(2008{\natexlab{a}})\citenamefont{Isaac,
  G\'omez~Rivas, Sambles, Barnes, and Hendry}}]{isaac:113411}
\bibinfo{author}{\bibfnamefont{T.~H.} \bibnamefont{Isaac}},
  \bibinfo{author}{\bibfnamefont{J.}~\bibnamefont{G\'omez~Rivas}},
  \bibinfo{author}{\bibfnamefont{J.~R.} \bibnamefont{Sambles}},
  \bibinfo{author}{\bibfnamefont{W.~L.} \bibnamefont{Barnes}},
  \bibnamefont{and} \bibinfo{author}{\bibfnamefont{E.}~\bibnamefont{Hendry}},
  \bibinfo{journal}{Phys. Rev. B} \textbf{\bibinfo{volume}{77}},
  \bibinfo{pages}{113411} (\bibinfo{year}{2008}{\natexlab{a}}),
  \urlprefix\url{http://link.aps.org/doi/10.1103/PhysRevB.77.113411}.

\bibitem[{\citenamefont{Zhu et~al.}(2011)\citenamefont{Zhu, Han, Tian, Gu,
  Chen, and Zhang}}]{zhu:3129}
\bibinfo{author}{\bibfnamefont{J.}~\bibnamefont{Zhu}},
  \bibinfo{author}{\bibfnamefont{J.}~\bibnamefont{Han}},
  \bibinfo{author}{\bibfnamefont{Z.}~\bibnamefont{Tian}},
  \bibinfo{author}{\bibfnamefont{J.}~\bibnamefont{Gu}},
  \bibinfo{author}{\bibfnamefont{Z.}~\bibnamefont{Chen}}, \bibnamefont{and}
  \bibinfo{author}{\bibfnamefont{W.}~\bibnamefont{Zhang}},
  \bibinfo{journal}{Optics Communications} \textbf{\bibinfo{volume}{284}},
  \bibinfo{pages}{3129 } (\bibinfo{year}{2011}), ISSN
  \bibinfo{issn}{0030-4018},
  \urlprefix\url{http://www.sciencedirect.com/science/article/pii/S0030401811002057}.

\bibitem[{\citenamefont{Hanham et~al.}(2012)\citenamefont{Hanham,
  Fernández-Domínguez, Teng, Ang, Lim, Yoon, Ngo, Klein, Pendry, and
  Maier}}]{hanham:226}
\bibinfo{author}{\bibfnamefont{S.~M.} \bibnamefont{Hanham}},
  \bibinfo{author}{\bibfnamefont{A.~I.} \bibnamefont{Fernández-Domínguez}},
  \bibinfo{author}{\bibfnamefont{J.~H.} \bibnamefont{Teng}},
  \bibinfo{author}{\bibfnamefont{S.~S.} \bibnamefont{Ang}},
  \bibinfo{author}{\bibfnamefont{K.~P.} \bibnamefont{Lim}},
  \bibinfo{author}{\bibfnamefont{S.~F.} \bibnamefont{Yoon}},
  \bibinfo{author}{\bibfnamefont{C.~Y.} \bibnamefont{Ngo}},
  \bibinfo{author}{\bibfnamefont{N.}~\bibnamefont{Klein}},
  \bibinfo{author}{\bibfnamefont{J.~B.} \bibnamefont{Pendry}},
  \bibnamefont{and} \bibinfo{author}{\bibfnamefont{S.~A.} \bibnamefont{Maier}},
  \bibinfo{journal}{Advanced Materials} \textbf{\bibinfo{volume}{24}},
  \bibinfo{pages}{OP226} (\bibinfo{year}{2012}), ISSN
  \bibinfo{issn}{1521-4095},
  \urlprefix\url{http://dx.doi.org/10.1002/adma.201202003}.

\bibitem[{\citenamefont{Jung et~al.}(2015)\citenamefont{Jung, Mandel, Bendoym,
  Golovin, and Crouse}}]{jung:1007}
\bibinfo{author}{\bibfnamefont{Y.~U.} \bibnamefont{Jung}},
  \bibinfo{author}{\bibfnamefont{I.~M.} \bibnamefont{Mandel}},
  \bibinfo{author}{\bibfnamefont{I.}~\bibnamefont{Bendoym}},
  \bibinfo{author}{\bibfnamefont{A.~B.} \bibnamefont{Golovin}},
  \bibnamefont{and} \bibinfo{author}{\bibfnamefont{D.~T.}
  \bibnamefont{Crouse}}, \bibinfo{journal}{J. Opt. Soc. Am. B}
  \textbf{\bibinfo{volume}{32}}, \bibinfo{pages}{1007} (\bibinfo{year}{2015}),
  \urlprefix\url{http://josab.osa.org/abstract.cfm?URI=josab-32-5-1007}.

\bibitem[{\citenamefont{Deng et~al.}(2013)\citenamefont{Deng, Teng, Liu, Wu,
  Tang, Zhang, Maier, Lim, Ngo, Yoon et~al.}}]{deng:128}
\bibinfo{author}{\bibfnamefont{L.}~\bibnamefont{Deng}},
  \bibinfo{author}{\bibfnamefont{J.}~\bibnamefont{Teng}},
  \bibinfo{author}{\bibfnamefont{H.}~\bibnamefont{Liu}},
  \bibinfo{author}{\bibfnamefont{Q.~Y.} \bibnamefont{Wu}},
  \bibinfo{author}{\bibfnamefont{J.}~\bibnamefont{Tang}},
  \bibinfo{author}{\bibfnamefont{X.}~\bibnamefont{Zhang}},
  \bibinfo{author}{\bibfnamefont{S.~A.} \bibnamefont{Maier}},
  \bibinfo{author}{\bibfnamefont{K.~P.} \bibnamefont{Lim}},
  \bibinfo{author}{\bibfnamefont{C.~Y.} \bibnamefont{Ngo}},
  \bibinfo{author}{\bibfnamefont{S.~F.} \bibnamefont{Yoon}},
  \bibnamefont{et~al.}, \bibinfo{journal}{Advanced Optical Materials}
  \textbf{\bibinfo{volume}{1}}, \bibinfo{pages}{128} (\bibinfo{year}{2013}),
  ISSN \bibinfo{issn}{2195-1071},
  \urlprefix\url{http://dx.doi.org/10.1002/adom.201200032}.

\bibitem[{\citenamefont{Isaac et~al.}(2008{\natexlab{b}})\citenamefont{Isaac,
  Barnes, and Hendry}}]{isaac:241115}
\bibinfo{author}{\bibfnamefont{T.~H.} \bibnamefont{Isaac}},
  \bibinfo{author}{\bibfnamefont{W.~L.} \bibnamefont{Barnes}},
  \bibnamefont{and} \bibinfo{author}{\bibfnamefont{E.}~\bibnamefont{Hendry}},
  \bibinfo{journal}{Applied Physics Letters} \textbf{\bibinfo{volume}{93}},
  \bibinfo{eid}{241115} (\bibinfo{year}{2008}{\natexlab{b}}),
  \urlprefix\url{http://scitation.aip.org/content/aip/journal/apl/93/24/10.1063/1.3049350}.

\bibitem[{\citenamefont{Nagel et~al.}(2006)\citenamefont{Nagel, Forst, and
  Kurz}}]{nagel:s601}
\bibinfo{author}{\bibfnamefont{M.}~\bibnamefont{Nagel}},
  \bibinfo{author}{\bibfnamefont{M.}~\bibnamefont{Forst}}, \bibnamefont{and}
  \bibinfo{author}{\bibfnamefont{H.}~\bibnamefont{Kurz}},
  \bibinfo{journal}{Journal of Physics: Condensed Matter}
  \textbf{\bibinfo{volume}{18}}, \bibinfo{pages}{S601} (\bibinfo{year}{2006}),
  \urlprefix\url{http://stacks.iop.org/0953-8984/18/i=18/a=S07}.

\bibitem[{\citenamefont{Ng et~al.}(2013)\citenamefont{Ng, Wu, Hanham,
  Fernández-Domínguez, Klein, Liew, Breese, Hong, and Maier}}]{ng:2195}
\bibinfo{author}{\bibfnamefont{B.}~\bibnamefont{Ng}},
  \bibinfo{author}{\bibfnamefont{J.}~\bibnamefont{Wu}},
  \bibinfo{author}{\bibfnamefont{S.~M.} \bibnamefont{Hanham}},
  \bibinfo{author}{\bibfnamefont{A.~I.} \bibnamefont{Fernández-Domínguez}},
  \bibinfo{author}{\bibfnamefont{N.}~\bibnamefont{Klein}},
  \bibinfo{author}{\bibfnamefont{Y.~F.} \bibnamefont{Liew}},
  \bibinfo{author}{\bibfnamefont{M.~B.~H.} \bibnamefont{Breese}},
  \bibinfo{author}{\bibfnamefont{M.}~\bibnamefont{Hong}}, \bibnamefont{and}
  \bibinfo{author}{\bibfnamefont{S.~A.} \bibnamefont{Maier}},
  \bibinfo{journal}{Advanced Optical Materials} \textbf{\bibinfo{volume}{1}},
  \bibinfo{pages}{543} (\bibinfo{year}{2013}), ISSN \bibinfo{issn}{2195-1071},
  \urlprefix\url{http://dx.doi.org/10.1002/adom.201300146}.

\bibitem[{\citenamefont{Melinger et~al.}(2009)\citenamefont{Melinger, Harsha,
  Laman, and Grischkowsky}}]{melinger:a79}
\bibinfo{author}{\bibfnamefont{J.~S.} \bibnamefont{Melinger}},
  \bibinfo{author}{\bibfnamefont{S.~S.} \bibnamefont{Harsha}},
  \bibinfo{author}{\bibfnamefont{N.}~\bibnamefont{Laman}}, \bibnamefont{and}
  \bibinfo{author}{\bibfnamefont{D.}~\bibnamefont{Grischkowsky}},
  \bibinfo{journal}{J. Opt. Soc. Am. B} \textbf{\bibinfo{volume}{26}},
  \bibinfo{pages}{A79} (\bibinfo{year}{2009}),
  \urlprefix\url{http://josab.osa.org/abstract.cfm?URI=josab-26-9-A79}.

\bibitem[{\citenamefont{Brown et~al.}(2007)\citenamefont{Brown, Bjarnason,
  Fedor, and Korter}}]{brown:061908}
\bibinfo{author}{\bibfnamefont{E.~R.} \bibnamefont{Brown}},
  \bibinfo{author}{\bibfnamefont{J.~E.} \bibnamefont{Bjarnason}},
  \bibinfo{author}{\bibfnamefont{A.~M.} \bibnamefont{Fedor}}, \bibnamefont{and}
  \bibinfo{author}{\bibfnamefont{T.~M.} \bibnamefont{Korter}},
  \bibinfo{journal}{Applied Physics Letters} \textbf{\bibinfo{volume}{90}},
  \bibinfo{eid}{061908} (\bibinfo{year}{2007}),
  \urlprefix\url{http://scitation.aip.org/content/aip/journal/apl/90/6/10.1063/1.2437107}.

\bibitem[{\citenamefont{Walther et~al.}(2005)\citenamefont{Walther, Freeman,
  and Hegmann}}]{walther:261107}
\bibinfo{author}{\bibfnamefont{M.}~\bibnamefont{Walther}},
  \bibinfo{author}{\bibfnamefont{M.~R.} \bibnamefont{Freeman}},
  \bibnamefont{and} \bibinfo{author}{\bibfnamefont{F.~A.}
  \bibnamefont{Hegmann}}, \bibinfo{journal}{Applied Physics Letters}
  \textbf{\bibinfo{volume}{87}}, \bibinfo{eid}{261107} (\bibinfo{year}{2005}),
  \urlprefix\url{http://scitation.aip.org/content/aip/journal/apl/87/26/10.1063/1.2158025}.

\end{thebibliography}

\end{document}